\documentclass[superscriptaddress,twocolumn,showpacs,prb]{revtex4-1}
\usepackage[utf8]{inputenc}
\usepackage{amsmath}
\usepackage{braket}
\usepackage{graphicx}
\usepackage{amsfonts}
\usepackage{pgfplots}
\usepackage{csquotes}
\usepackage{hhline}
\usepackage{amssymb}
\usepackage{listings}
\usepackage{color}

\definecolor{codegreen}{rgb}{0,0.6,0}
\definecolor{codegray}{rgb}{0.5,0.5,0.5}
\definecolor{codepurple}{rgb}{0.58,0,0.82}
\definecolor{backcolour}{rgb}{0.95,0.95,0.92}
 
\lstdefinestyle{mystyle}{
    backgroundcolor=\color{backcolour},   
    commentstyle=\color{codegreen},
    keywordstyle=\color{magenta},
    numberstyle=\tiny\color{codegray},
    stringstyle=\color{codepurple},
    basicstyle=\footnotesize,
    breakatwhitespace=false,         
    breaklines=true,                 
    captionpos=b,                    
    keepspaces=true,                 
    numbers=left,                    
    numbersep=5pt,                  
    showspaces=false,                
    showstringspaces=false,
    showtabs=false,                  
    tabsize=2
}
 
\usepackage{subfigure}

\usepackage{dcolumn}
\usepackage{tabularx}
\usepackage[colorlinks=true,linkcolor=blue,citecolor=blue,urlcolor=blue]{hyperref}
\usepackage{longtable}
\usepackage{braket}
\usepackage{float}
\newcolumntype{C}{>{\centering\arraybackslash}X}
\begin{document}
\title{Solving Vehicle Routing Problem Using Quantum Approximate Optimization Algorithm}

\author{Utkarsh}
\email{utkarsh.azad@research.iiit.ac.in}
\affiliation{Center for Computational Natural Sciences and Bioinformatics, \\
International Institute of Information Technology Hyderabad, Hyderabad 500032, Telangana, India}
\author{Bikash K. Behera}
\email{bikas.riki@gmail.com}
\affiliation{Bikash's Quantum (OPC) Pvt. Ltd.,\\ Balindi, Mohanpur 741246, Nadia, West Bengal, India}
\affiliation{Department of Physical Sciences,\\ Indian Institute of Science Education and Research Kolkata, Mohanpur 741246, West Bengal, India}
\author{Emad A. Ahmed}
\email{emad.amer@sci.svu.edu.eg}
\affiliation{Department of Computer Science, Faculty of Computers and Information, South Valley University, Qena, Egypt}
\author{Prasanta K. Panigrahi}
\email{pprasanta@iiserkol.ac.in}
\affiliation{Department of Physical Sciences,\\ Indian Institute of Science Education and Research Kolkata, Mohanpur 741246, West Bengal, India}
\author{Ahmed Farouk}
\email{afarouk@wlu.ca}
\affiliation{Department of Computer Science, Faculty of Computers and Artificial Intelligence, South Valley University, Hurghada, Egypt}

\begin{abstract}
In this paper, we describe the usage of the Quantum Approximate Optimization Algorithm (QAOA), which is a quantum-classical heuristic, to solve a combinatorial optimization and integer programming task known as Vehicle Routing Problem (VRP). We outline the Ising formulation for VRP and present a detailed procedure to solve VRP by minimizing its simulated Ising Hamiltonian using the IBM Qiskit platform. Here, we attempt to find solutions for the VRP problems: $(4,2)$, $(5,2)$, and $(5,3)$, where each (n, k) represents a VRP problem with $n$ locations and $k$ vehicles. We find that the performance of QAOA is not just dependent upon the classical optimizer used, the number of steps \textit{p} in which an adiabatic path is realized, or the way parameters are initialized, but also on the problem instance itself.
\end{abstract}

\maketitle

\section{Introduction}
A majority of the real world applications involve mapping the task at hand to an optimization problem, whose solutions are either fully known or can be approximated by relaxing some of its constraints. In general, the increment in problem size or augmentation of additional constraints can increase the hardness of the problem. This means that the computational resources required in solving them scale-up exponentially, making them computationally intractable. Hence, researchers worldwide have been working on developing efficient tools and techniques for solving these problems efficiently with the available computing power. 

With the recent development of quantum processors by IBM \cite{bib_IBM}, Rigetti \cite{bib_Pyquil}, Google \cite{bib_Google} etc, various proposals have been made to use them to find solutions for optimization problems. In general, quantum computing devices are supposed to have a computational advantage over classical processors by using quantum resources such as superposition and entanglement. However, the computational capabilities of these current generation quantum processors also known as Noisy Intermediate-Scale Quantum (NISQ) \cite{bib_Preskill} devices, are considerably restricted due to their intermediate size (in terms of qubits count), limited connectivity, imperfect qubit-control, short coherence time and minimal error-correction. Hence, they are only able to run algorithms with limited circuit depth. The Quantum Approximate Optimization Algorithm (QAOA) \cite{bib_Farhi} is one of such algorithm which belongs to the class of quantum-classical hybrid variational algorithms. It can be thought of as a coarsely trotterized adiabatic time evolution in $p$ steps to $\Ket{\psi^{H_c}_{GS}}$ i.e., the ground state of a Hamiltonian $H_c$ which encodes the problem from $\Ket{\psi^{H_m}_{GS}}$ i.e., the ground state of the Hamiltonian $H_m$ which is known and easier to prepare. Thus, it can be used to solve combinatorial optimization problems mapped to the minimization of an Ising Hamiltonian on near-term devices. As the Ising problem is itself NP-hard, QAOA is an expected candidate for demonstrating quantum supremacy. Here, in this paper, we use QAOA to solve the Vehicle Routing Problem (VRP), which is an NP-hard combinatorial optimization problem \cite{bib_Feld}.

\textit{Structure} -- In Section \ref{VRP}, we discuss VRP and present its Ising formulation in Section \ref{IFVRP}. Then, in Section \ref{QAOA}, we explain QAOA in detail and present our simulation results in Section \ref{SimRes}. Finally, in Section \ref{Diss}, we discuss the performance and limitations of using QAOA to solve combinatorial optimizations problem in general.

\section{Vehicle Routing Problem} \label{VRP}
Vehicle Routing Problem is an NP-hard combinatorial optimization problem. Any problem instance $(n,k)$ of VRP involves $k$ vehicles, and $n-1$ locations (other than the depot $D$). Its solution is the set of routes in which all of the $k$ vehicles begin and end in the $D$, such that each location is visited exactly once. The optimal route is the one in which the total distance travelled by $k$ vehicles is least. In a way, this problem is a generalization of the classic Travelling Salesman Problem \cite{bib_TSP_Bik}, where now a group of $k$ salesmen begin and end at the same location $D$ while having to collectively serve $n-1$ locations such that each location is served exactly once.

In most real world applications, the VRP problem \cite{bib_VRP_Bik} is generally augmented by constraints, such as vehicle capacity or limited coverage-time. However, here we only focus on showing how to solve the most basic version of VRP without any of these additional constraints.

\section{Ising Formulation of VRP} \label{IFVRP}
To solve a problem instance $(n,k)$ of VRP using QAOA, we first need to map it to the minimization of an Ising Hamiltonian $H_c$ \cite{bib_Lucas}. We do this by first finding the $H_c$, which encodes the given problem instance.

Let $x_{ij}$ be the binary decision variable which has the value $1$ if there exists an edge from $i$ to $j$ with weight $w_{ij} > 0$, else it is $0$. To represent a solution to VRP problem, there will be $n\times (n-1)$ decision variables. Next, we for every edge $i\rightarrow j$ we define two sets $source [i]$ and $target [j]$. The set $source [i]$ will contain the nodes $j$ to which node $i$ sends an edge. Similarly, the set $target [j]$ will contain all the nodes $i$ which send an edge to node $j$. Hence, the VRP can be forumalted as 
\begin{equation}\label{eq:1}
    VRP(n,k) = min_{\{x_{ij}\}_{i\rightarrow j} \in \{0,1\}} \sum_{i\rightarrow j} w_{ij}x_{ij}
\end{equation}
Subjected to the following constraints: 
\begin{equation}\label{eq:2}
	\sum_{j \in source[i]} x_{ij} = 1 \quad\forall i \in \{1, \ldots, n-1\}
\end{equation}
\begin{equation}\label{eq:3}
	\sum_{j \in target[i]} x_{ji} = 1 \quad\forall i \in \{1, \ldots, n-1\}
\end{equation}
\begin{equation}\label{eq:4}
	\sum_{j \in source[0]} x_{0j} = k
\end{equation}
\begin{equation}\label{eq:5}
	\sum_{j \in target[0]} x_{j0} = k
\end{equation}

Here, Eqs. (\ref{eq:2}), (\ref{eq:3}) impose the node-visiting constraint so that each node is visited exactly once. Also, the Eqs. (\ref{eq:4}), (\ref{eq:5}) impose the constraint to enforce that all the vehicles begin from and return back to depot $D$, i.e., the node $0$. Now, using Eqs. (\ref{eq:1}-\ref{eq:5}), the energy functional $H_{VRP}$ of the above problem can be written as:
\begin{equation}\label{eq:6}
    H_{VRP} = H_A + H_B + H_C + H_D + H_E
\end{equation}
\begin{equation}\label{eq:7}
    H_{A} = \sum_{i\rightarrow j} w_{ij}x_{ij}
\end{equation}
\begin{equation}\label{eq:8}
    H_{B} = A\sum_{i \in 1,\ldots,n-1} \Big(1 - \sum_{j \in source[i]} x_{ij} \Big)^2
\end{equation}
\begin{equation}\label{eq:9}
    H_{C} = A\sum_{i \in 1,\ldots,n-1} \Big(1 - \sum_{j \in target[i]} x_{ji} \Big)^2
\end{equation}
\begin{equation}\label{eq:10}
    H_{D} = A \Big(k - \sum_{j \in source[0]} x_{0j} \Big)^2
\end{equation}
\begin{equation}\label{eq:11}
    H_{E} = A \Big(k - \sum_{j \in target[0]} x_{j0} \Big)^2
\end{equation}

Here, $A>0$ is a constant, which is dependent on the problem instance itself. Next, for $VRP (n,k)$, we can represent all the decision variables $x_{ij}$ using the following vector $\vec{x}$:
\begin{equation}\label{eq:12}
    \vec{\textbf{x}} = [x_{(0,1)}, x_{(0,2)}, \ldots x_{(1,0)}, x_{(1,2)}, \ldots x_{(n-1,n-2)}]^{\textbf{T}}
\end{equation}

Using \textbf{$\vec{x}$}, we also define the following vectors for every node $i$: $\vec{z}_{S[i]}$ and $\vec{z}_{T[i]}$. The first vector $\vec{z}_{S[i]}$ is $\vec{x}$ with $x_{ij}=1$, $x_{kj}=0$ if $k \neq i$,  $\forall j,k \in \{0,\ldots,n-1\}$. Similarly, the second vector $\vec{z}_{T[i]}$ is $\vec{x}$ with $x_{ji}=1$, $x_{jk}=0$ if $k \neq i$, $\forall j,k \in \{0,\ldots,n-1\}$. These two additional vectors could be understood more clearly using the following two Eqs. (\ref{eq:add1}-\ref{eq:add2}):

\begin{equation}\label{eq:add1}
    \sum_{j \in source[i]} x_{ij} = \vec{z}_{S[i]}^{\textbf{T}}\vec{x}
\end{equation}
\begin{equation}\label{eq:add2}
    \sum_{j \in target[i]} x_{ji} = \vec{z}_{T[i]}^{\textbf{T}}\vec{x}
\end{equation}

From Eq. (\ref{eq:12}), we can represent minimization of $H_{VRP}$ as a quadratic unconstrained binary optimization (QUBO) problem which is isomorphic to the ising problem. In general, a QUBO problem for a graph $G = (N,V)$ may be defined as: 
\begin{equation}\label{eq:13}
    f(x)_{QUBO} = min_{x \in \{0,1\}^{(N\times V)}}\enspace x^{T}Qx + g^{T}x + c 
\end{equation}
Here the quadratic coefficient $Q$ represents the edge weight i.e., coupling or interaction between two nodes, the linear coefficient $g$ represents the node weight i.e., contribution from individual nodes, and the term $c$ is a constant offset. In order to find these coefficient in the QUBO forumations  \cite{bib_DWAVE} of $H_{VRP}$ given in Eq. (\ref{eq:6}) we first put in Eqs. (\ref{eq:add1}-\ref{eq:add2}) in Eqs. (\ref{eq:8}-\ref{eq:9}) respectively, then expand and regroup Eq. (\ref{eq:6}) according to Eq. (\ref{eq:13}). 
\begin{equation} \label{eq:add3}
    \begin{split}
     H & = A\sum_{i=0}^{n-1}[z_{S[i]}z_{S[i]}^{T} + z_{T[i]}z_{T[i]}^{T}]\vec{x}^{2} + \\
     & w^{T}\vec{x} - 2A\sum_{i=1}^{n-1}[z_{S[i]}^{T} + z_{T[i]}^{T}]\vec{x} - \\
     & 2Ak[z_{S[0]}^{T} + z_{T[0]}^{T}]\vec{x} + 2A(n-1) + 2Ak^{2}
    \end{split}
\end{equation}

\begin{figure}[ht]
    \centering
    \includegraphics[scale=0.6]{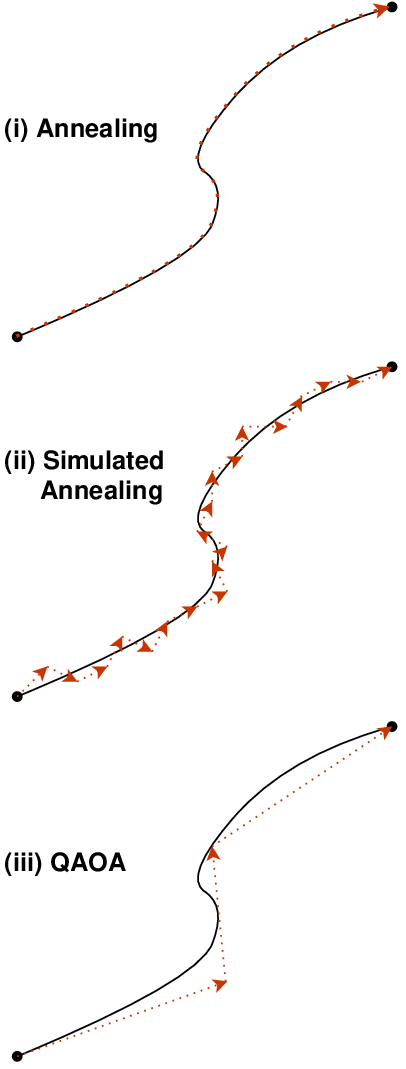}
    \caption{Consider an adiabatic time evolution path (black) in the state space. Conceptually, quantum annealing (top) follows this adiabatic time evolution path. Simulated annealing (middle) follows this path in discrete steps i.e. it follows a trotterized adiabatic time evolution path. QAOA follows this path in $p$ steps, i.e. it follows a coarsely trotterized adiabatic time evolution. \cite{bib_Verdon}}
    \label{fig:Fig1}
\end{figure}

From Eq. (\ref{eq:add3}), we get the coefficients Q \textit{$(n(n-1)\times n(n-1))$}, g \textit{$(n(n-1)\times 1)$} and c:

\begin{equation}\label{eq:14}
    \begin{split}
    Q & = A \Big[[z_{T[1]}, z_{T[2]}, \ldots, z_{T[n-1]}, z_{T[0]}, z_{T[2]}, \ldots, z_{T[n-2]}]^T,  \\
      & \quad + [[Z_{S[0]}]^{\times(n-1)}
      [Z_{S[1]}]^{\times(n-1)}  \ldots [Z_{S[n-1]}]^{\times(n-1)}]\Big]
    \end{split}
\end{equation}
\begin{equation}\label{eq:15}
    \begin{split}
    g & = w - 2A[J + K] - 2Ak[z_{S[0]} + z_{T[0]}] \\
    \end{split}
\end{equation}
\begin{equation}\label{eq:16}
    c = 2A(n - 1) + 2Ak^2
\end{equation}

Here, $J$ is a $n\times(n-1)$ vector with first $n-1$ elements $0$ and rest $(n-1)^{2}$ elements $1$, vector $K$ is $\vec{x}$ with $x_{ij}=1$ if $j \neq 0$, $\forall i \in \{0,\ldots,n-1\}$, else $0$ and $\vec{w}$ is a weight vector. From this, to construct Ising Hamiltonian for $VRP$ we expand Eq. (\ref{eq:13}) by using Eqs. (\ref{eq:14}-\ref{eq:16}) and rewrite all the binary variables $x_{ij} \in \{0,1\}$ using spin variables $s_{ij} \in \{-1,1\}$.

\begin{equation}\label{eq:17}
    x_{ij} = \frac{s_{ij} + 1}{2}
\end{equation}

By regrouping \cite{bib_DWAVE} together similar terms from the expansion, one can express $\hat{H}_{ISING}$ for $VRP(n,k)$ as:
\\ 
\\
\begin{equation}\label{eq:18}
    H_{ISING} = -\sum_{i}\sum_{j<i} J_{ij}s_{i}s_{j} + \sum_{i}h_{i}s_{i} + d
\end{equation}
\begin{equation}\label{eq:19}
   J_{ij} = -\frac{Q_{ij}}{4} \quad \forall i < j, \quad J_{ii} = 0 \quad \forall i
\end{equation}
\begin{equation}\label{eq:20}
    h_{i} = \frac{g_{i}}{2} + \sum_{j}\frac{Q_{ij}}{4} + \sum_{j}\frac{Q_{ji}}{4}
\end{equation}
\begin{equation}\label{eq:21}
    d = c + \sum_{i} \frac{g_{i}}{2} + \sum_{i} \frac{Q_{ii}}{4} + \sum_{i}\sum_{j}\frac{Q_{ij}}{4} 
\end{equation}

Replacing $s_{i}$ with $\sigma_{i}^{z}$ i.e. the Pauli-Z operator acting on $i^{th}$ qubit, gives the quantum mechanical description of $\hat{H}_{ISING}$ which is implementable on a quantum computer. 

\section{Quantum Approximate Optimization Algorithm} \label{QAOA}
Adiabatic quantum computation (AQC)\cite{bib_Albash} was the first quantum computation model to be used for solving hard combinatorial optimization problems. Unlike the gate-based quantum computation model, it was based on adiabatic theorem from quantum mechanics. In this model, to perform any computation we need two Hamiltonians called $\hat{H}_{mixer}$ and $\hat{H}_{cost}$. Amongst them, the ground state of $\hat{H}_{mixer}$ should be an easily preparable state such as $\Ket{+}^{\otimes N}$ and ground state of $\hat{H}_{cost}$ encode the solution to our problem. Both Hamiltonians $\hat{H}_{mixer}$ and $\hat{H}_{cost}$, should be local, i.e. they only involve terms for interactions between a constant number of particles. The instantaneous Hamiltonian $\hat{H}(t)$ for the system is:

\begin{equation}\label{eq:22}
    \hat{H}(t) = (1-t)\hat{H}_{mixer} + t \hat{H}_{cost}
\end{equation}

In AQC, $\Delta E$, i.e., the difference between the ground state and first excited state energy of $\hat{H}(t)$ bounds the step size one can take to follow adiabatic pathway \cite{bib_Ahar}. Hence, the computation time to solve any problem rises exponentially as $\Delta E$ becomes infinitesimally small. This limits its capability to solve a certain instance of hard optimization problems.

In Quantum Approximate Optimization Algorithm (QAOA), we eliminate this restriction on the step size. Instead, whole of the adiabatic pathway is discretized in some $p$ steps, where $p$ represents precision (Fig. \ref{fig:Fig1}). To do this, we trotterize the unitary into $p$ steps using the parameters $\{\beta, \gamma\}$ as follows:

\begin{equation}\label{eq:23}
\begin{split}
    U = U(\hat{H}_{mixer}, \beta_0)U(\hat{H}_{cost}, \gamma_0) \ldots \\ U(\hat{H}_{mixer}, \beta_{p-1})U(\hat{H}_{cost}, \gamma_{p-1})    
\end{split}
\end{equation}

\begin{figure*}[ht]
    \centering
    \includegraphics[scale=0.4]{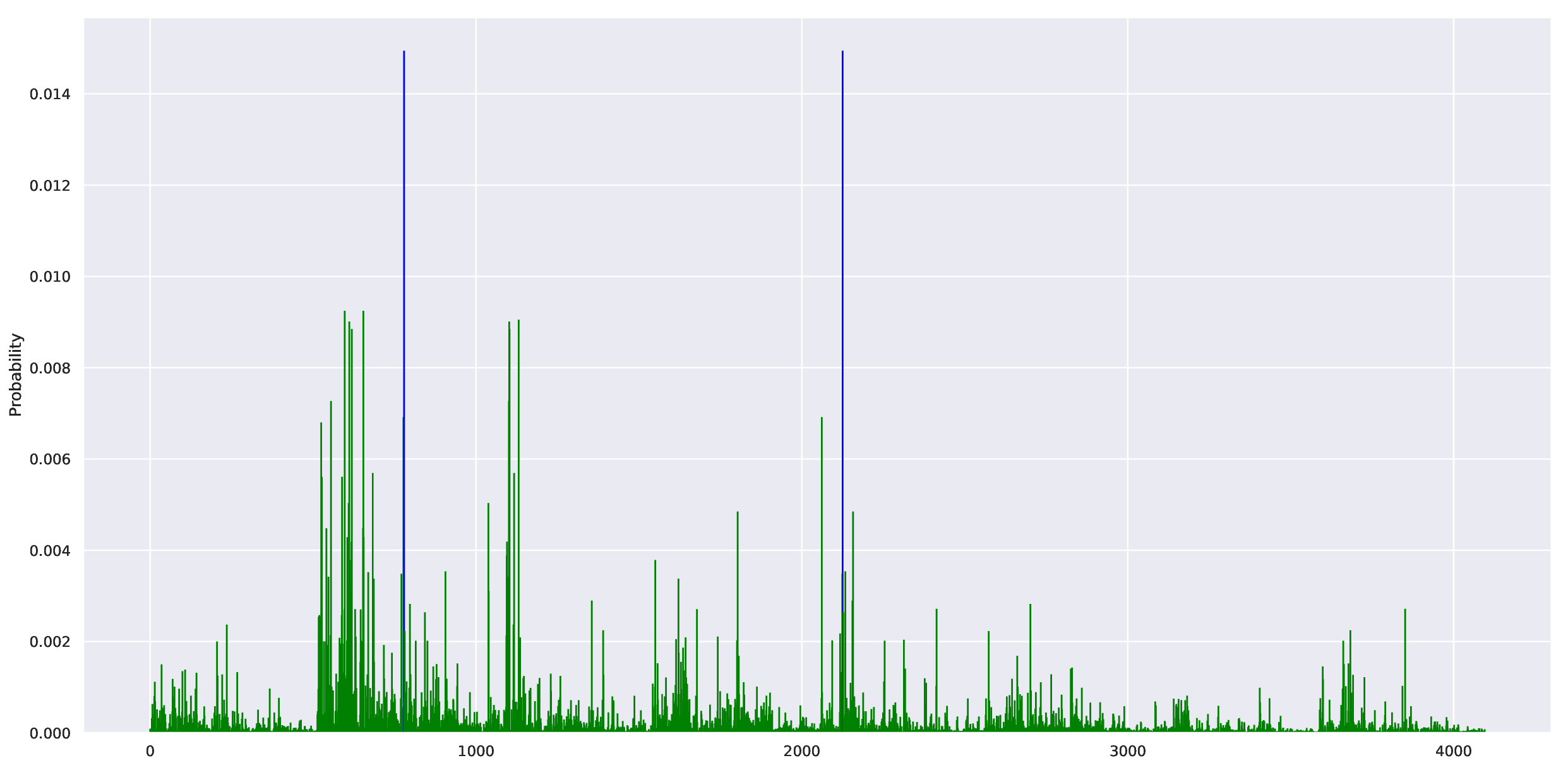}
    \caption{Probability distribution for the problem instance $(4,2)$ for $p=12$. Indices of optimal states (blue) are: $779$ and $2125$.}
    \label{fig:Fig2}
\end{figure*}

In gate-model quantum computation this means \cite{bib_Ahar} that starting from some initial product state $\ket{\psi^{HM}_{GS}}$, we apply a parameterized gate sequence to produce the state $\ket{\psi^{HC}}$. For some optimal value of the parameters: $\{\beta^{*}, \gamma^{*}\}$, this is the ground state of $\hat{H}_{cost}$. The parameters $\{\beta, \gamma\}$ are provided by a classical processor, and also optimized by a classical optimization routine based on the result of energy measurement for final state $\ket{\psi^{HC}}$. Hence, QAOA belongs to the class of hybrid quantum-classical variational algorithms.

\section{Simulations and Results} \label{SimRes}

We have executed QAOA using IBM Qiskit to solve VRP for three problem instances: $(4,2)$, $(5,2)$, and $(5,3)$, where each $(n,k)$ represents a problem with $n$ locations and $k$ vehicles with a distance matrix $D$ representing the squared euclidean distances between locations. One needs $N = n\times (n-1)$ qubits to encode the problem instance, i.e., state of each qubit represents the possibility of an edge between two nodes. The mixing Hamiltonian $\hat{H}_{mixer}$ and the cost Hamiltonian $\hat{H}_{cost}$ for this problem are:

\begin{equation}\label{eq:24}
    \hat{H}_{mixer} = -\sum^{n\times (n-1)-1}_{i=0} \sigma^{x}_{i} 
\end{equation}

\begin{equation}\label{eq:25}
  \hat{H}_{cost} = -\sum_{i}\sum_{j<i} J_{ij}\sigma^{z}_{i}\sigma^{z}_{j} - \sum_{i}h_{i}\sigma^{z}_{i} - d
\end{equation}

In $\hat{H}_{cost}$, we have $n\times (n-1) \times (n-2)$ terms of $J_{ij}$, $n\times (n-1)$ terms of $h_{i}$, and $d$ is an offset. In each of the following cases, we begin with the state $\ket{+}^{n\otimes (n-1)}$, which is the ground state of $\hat{H}_{mixer}$ given in Eq. (\ref{eq:24}). This state is prepared by applying Hadamard on all qubits initialized to $\ket{0}$. From Eq. (\ref{eq:23}), this state is evolved as:

\begin{equation}\label{eq:26}
    \begin{split}
    \ket{\beta, \gamma} = e^{-i \hat{H}_{mixer}\beta_p }e^{-i \hat{H}_{cost}\gamma_p }\ldots\\ e^{-i \hat{H}_{mixer}\beta_0 }e^{-i \hat{H}_{cost}\gamma_0 } \ket{+}^{n\otimes(n-1)}
    \end{split}
\end{equation}

For the evolved state $\ket{\beta,\gamma}$, we calculate energy $E$ by measuring the expectation value of $\hat{H}_{cost}$ as: 

\begin{equation}\label{eq:27}
    E = \bra{\beta,\gamma} \hat{H}_{cost} \ket{\beta,\gamma}
\end{equation}

Running a classical optimization routine on Eq. (\ref{eq:27}), we get the optimal value of $\{\beta,\gamma\}$ as $\{\beta^{*},\gamma^{*}\}$. To get the final result we measure the state $\ket{\beta^{*},\gamma^{*}}$. As shown in Fig. (\ref{fig:Fig2}), the state can collapse to any of the $2^{n\times (n-1)}$ possibilities. To visualize the solution, we represent the index of the collapsed state as $2^{n\times (n-1)}$ length bit string which represents flattened version of the adjacency matrix of the graph.

\subsection{Experiment 1} \label{exp:1}
In the first experiment, we have solved the problem instance $(4,2)$, described by the following distance matrix:

\begin{equation}
D_{1} = 
\begin{bmatrix}
0. & 36.84 &  5.06 & 30.63 \\
36.84 &  0. & 24.55 & 63.22 \\
5.06 & 24.55 & 0. & 15.50 \\
30.63 &  63.22 & 15.50 &  0 \\        
\end{bmatrix}    
\end{equation}

\begin{figure*}[htp]
    \centering
    \includegraphics[scale=0.38]{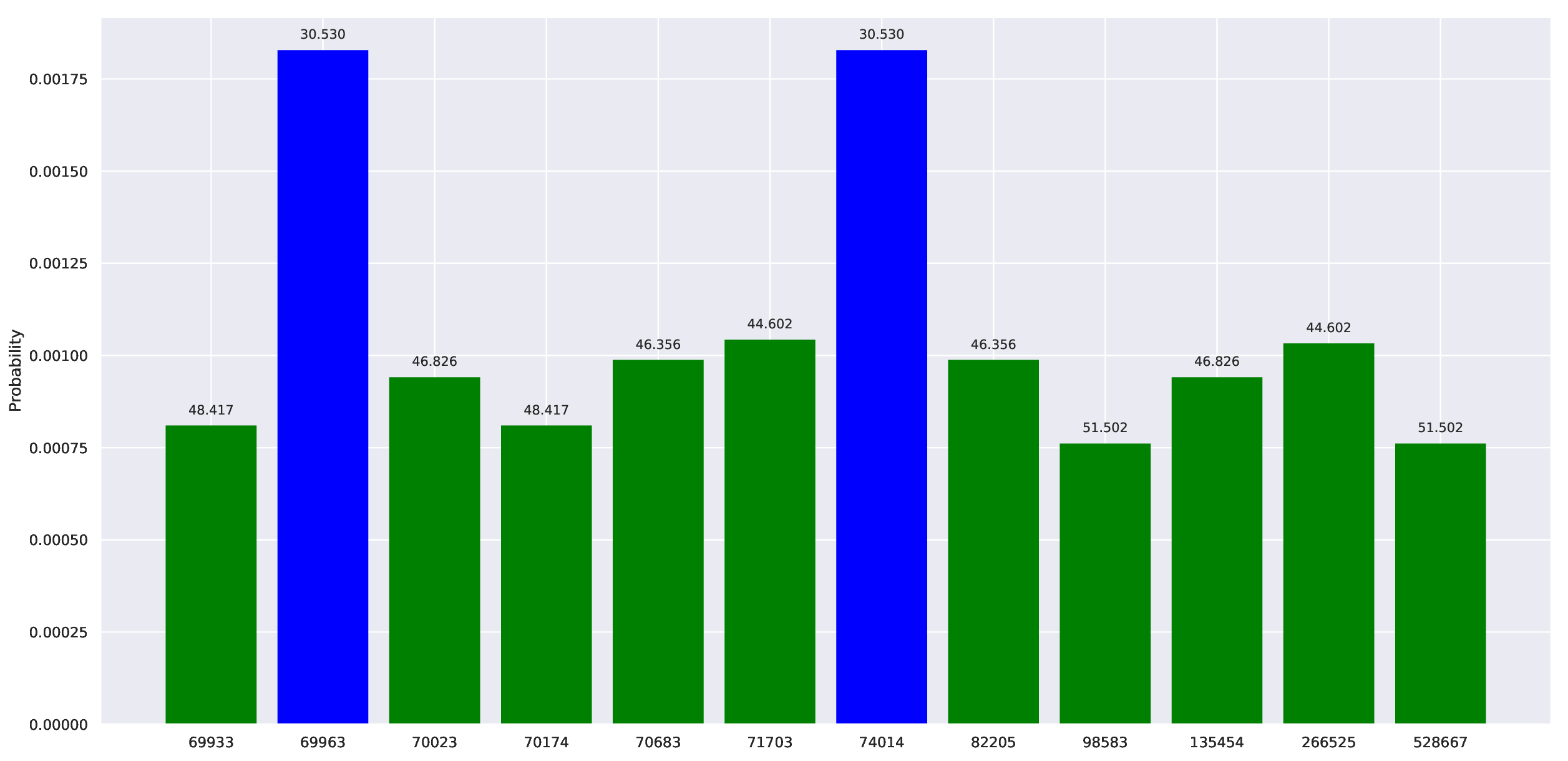}
    \caption{Probability distribution for top-12 feasible solutions of the problem instance $(5,3)$ for $p=24$. Indices of optimal states (blue) are: $69963$ and $74014$. Costs corresponding to each state are written at the top of the bars.}
    \label{fig:Fig6}
\end{figure*}

\begin{figure}[ht]
    \centering
    \includegraphics[width=0.48\textwidth]{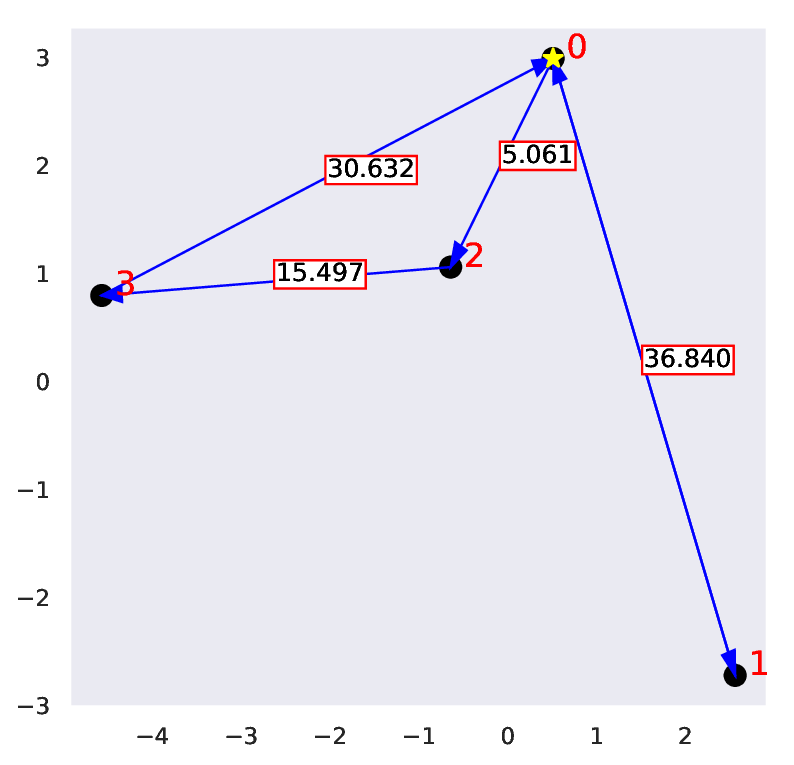}
    \caption{Visualization of the solution state indexed $779 = ``110100001100"$. The cost is: $C_{1} = 30.632 + 15.497 + 5.061 + 2\times 36.840 = 124.871$. Here, the node with yellow star denotes the depot, or the origin.}
    \label{fig:Fig3}
    \vspace*{-3pt}
\end{figure}

To encode the problem, we used $N=4\times3=12$ qubits. Using COBYLA optimizer, for $p\geq 12$ we were able to get the correct solution with sufficient probability. In Fig. (\ref{fig:Fig2}), we have shown the probability distribution of our result for $p=12$. States corresponding to indexes $779$ and $2125$ are equiprobable and solution states. We present the visualization of the state indexed $779 = [1, 1, 0, 1, 0, 0, 0, 0, 1, 1, 0, 0]$ in Fig. (\ref{fig:Fig3}) using the following adjacency matrix $A_1$:

\begin{equation}
A_{1} = 
\begin{bmatrix}
X & 1 & 1 & 0 \\
1 & X & 0 & 0 \\
0 & 0 & X & 1 \\
1 & 0 & 0 & X \\        
\end{bmatrix}    
\end{equation}

The cost in both cases come out to be $C_{1} = 30.632 + 15.497 + 5.061 + 2\times 36.840 = 124.871$.

\subsection{Experiment 2} \label{exp:2}

In the second experiment, we have solved the problem instance $(5,2)$, described by the following distance matrix:

\begin{equation}
D_{2} = 
\begin{bmatrix}

0 & 6.794 & 61.653 & 24.557 & 47.767 \\
6.794 &  0  & 87.312 & 47.262 & 39.477 \\
61.653 & 87.312 &  0.  &  9.711 & 42.887\\
24.557 & 47.262 &  9.711 &  0   & 40.98  \\
47.767 & 39.477 & 42.887 & 40.98  &  0 \\        
\end{bmatrix}    
\end{equation}

\begin{figure}[ht]
    \centering
    \includegraphics[width=0.48\textwidth]{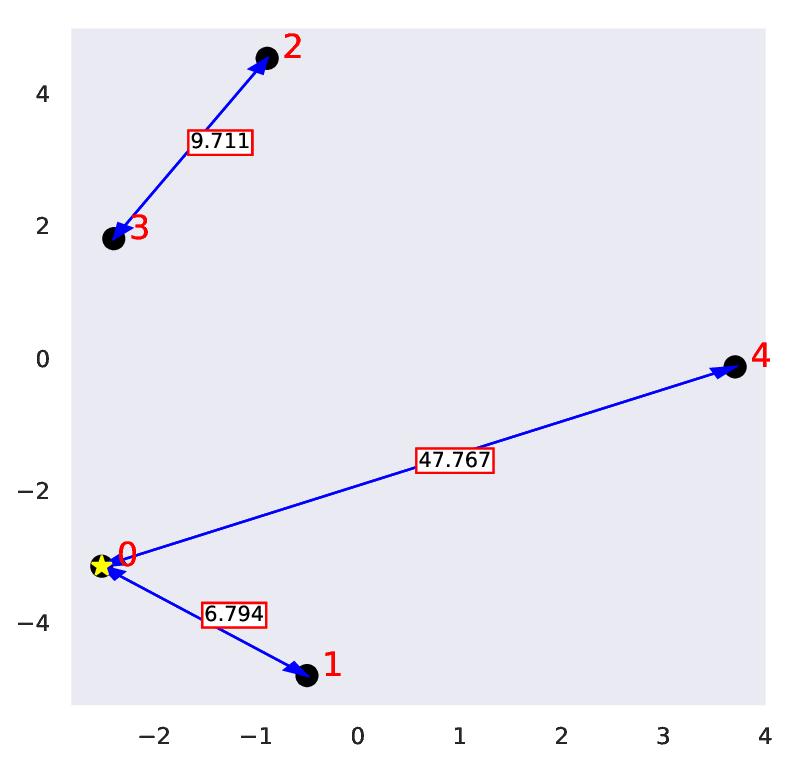}
    \caption{Visualization of the QAOA state indexed $623144 = ``10011000001000101000"$. The cost is: $C_{2} = 2\times 9.711 + 2\times 47.767 + 2\times 6.794 = 128.545$. Here, the node with yellow star denotes the depot, or the origin.}
    \label{fig:Fig4}
\end{figure}

To encode the problem, we used $N=5\times4=20$ qubits. We tried COBYLA, NELDER MEAD, and L-BFGS-B optimizers, for different values of $p$ ranging from $6$ to $40$. Fig. ($\ref{fig:Fig4}$) represents the visualization of the state indexed: $623144 = [1, 0, 0, 1, 1, 0, 0, 0, 0, 0, 1, 0, 0, 0, 1, 0, 1, 0, 0, 0]$, i.e., the state we got, using the following adjacency matrix $A_2$:

\begin{equation}
A_{2} = 
\begin{bmatrix}
X & 1 & 0 & 0 & 1 \\
1 & X & 0 & 0 & 0 \\
0 & 0 & X & 1 & 0 \\
0 & 0 & 1 & X & 0 \\
1 & 0 & 0 & 0 & X \\
\end{bmatrix}    
\end{equation}

\begin{figure}[ht]
    \centering
    \includegraphics[width=0.48\textwidth]{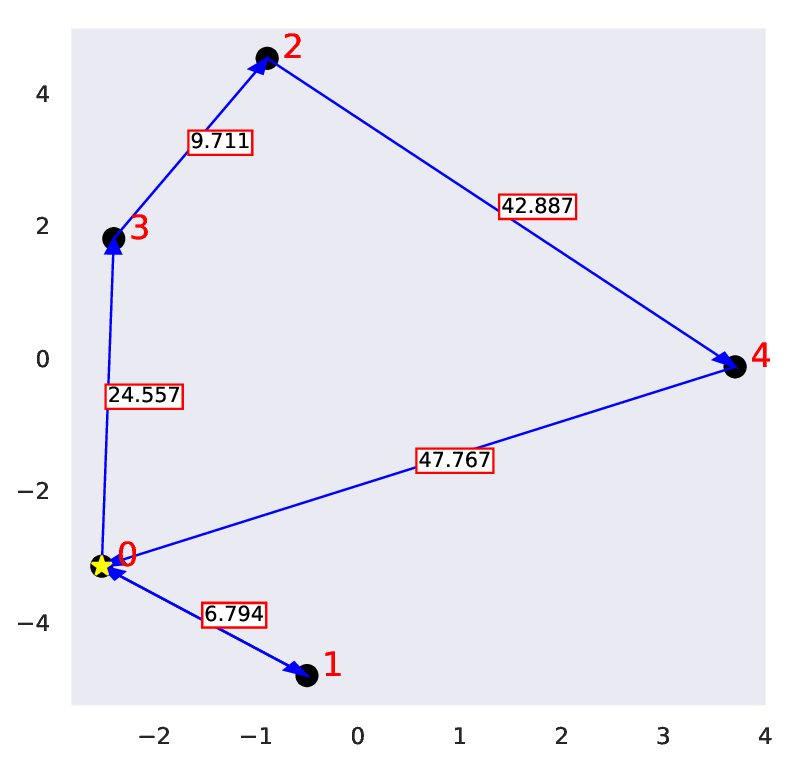}
    \caption{Visualization of the solution state indexed $688424 = ``10101000000100101000"$. The cost is: $C_{3} = 24.557 + 9.711 + 42.887 + 47.767 + 2\times 6.794 = 138.511$. Here, the node with yellow star denotes the depot, or the origin.}
    \label{fig:Fig5}
\end{figure}

The cost in this case comes out to be $C_{2} = 2\times 9.711 + 2\times 47.767 + 2\times 6.794 = 128.545$.

Where as, Fig. ($\ref{fig:Fig5}$) represents the visualization of the state indexed: \\$688424 = [1, 0, 1, 0, 1, 0, 0, 0, 0, 0, 0, 1, 0, 0, 1, 0, 1, 0, 0, 0]$, i.e., the optimal state we should have got, using the following adjacency matrix $A_3$:

\begin{equation}
A_{3} = 
\begin{bmatrix}
X & 1 & 0 & 1 & 0 \\
1 & X & 0 & 0 & 0 \\
0 & 0 & X & 0 & 1 \\
0 & 0 & 1 & X & 0 \\
1 & 0 & 0 & 0 & X \\
\end{bmatrix}   
\end{equation}

The cost in this case comes out to be $C_{3} = 24.557 + 9.711 + 42.887 + 47.767 + 2\times 6.794 = 138.511$. Therefore, as shown in Fig. ($\ref{fig:Fig4}$), the solution we got was not the optimal one i.e., given in Fig. $(\ref{fig:Fig5})$.

\subsection{Experiment 3} \label{exp:3}

In the third experiment, we have solved the problem instance $(5,3)$, described by the following distance matrix:

\begin{equation}
D_{4} = 
\begin{bmatrix}
0. & 12.138 & 0.32 & 7.2 & 2.626\\
12.138 & 0. & 16.307 & 5.3 & 17.021\\
0.32 & 16.307 &  0. & 9.309 & 2.98\\
7.2 & 5.3 & 9.309 & 0. & 16.759\\
2.626 & 17.021 & 2.98 & 16.759 &  0.\\        
\end{bmatrix}    
\end{equation}

To encode the problem, we used $N=5\times4=20$ qubits. Using COBYLA optimizer, for $p\geq 24$ we were able to get the correct solution with sufficient probability. In Fig. (\ref{fig:Fig6}), we have shown the probability distribution of our result for $p=24$ for top 12 feasible solutions. States corresponding to indexes $69963$ and $74014$ are equiprobable and solution states. 

\begin{figure}[ht]
    \centering
    \includegraphics[width=0.45\textwidth]{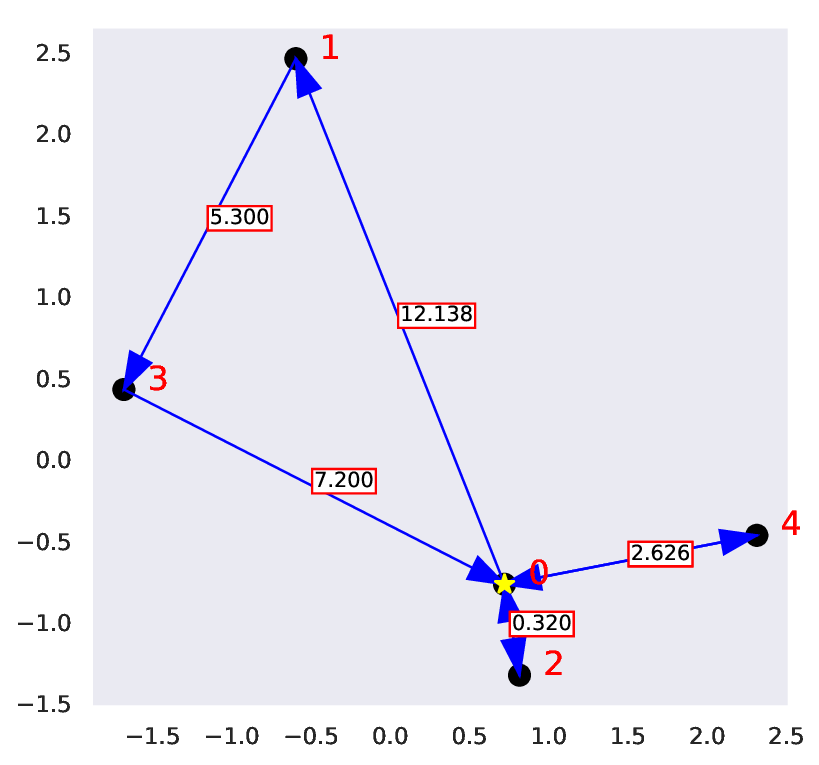}
    \caption{Visualization of the solution state indexed $69963 = ``10001000101001011"$. The cost is: $C_{4} = 12.138 + 5.300 + 7.200 + 2\times 2.626 + 2\times 0.320 = 30.530$. Here, the node with yellow star denotes the depot, or the origin.}
    \label{fig:Fig7}
\end{figure}

We present the visualization of the state $69963 = [1, 0, 0, 0, 1, 0, 0, 0, 1, 0, 1, 0, 0, 1, 0, 1, 1]$ in Fig. (\ref{fig:Fig7}) using the following adjacency matrix $A_4$:

\begin{equation}
A_{4} = 
\begin{bmatrix}
X & 1 & 1 & 0 & 1 \\
0 & X & 0 & 1 & 0 \\
1 & 0 & X & 0 & 0 \\
1 & 0 & 0 & X & 0 \\
1 & 0 & 0 & 0 & X \\
\end{bmatrix}    
\end{equation}

The cost in both cases come out to be $C_{4} = 12.138 + 5.300 + 7.200 + 2\times 2.626 + 2\times 0.320 = 30.530$.

\section{Discussions} \label{Diss}

Here, we have used Quantum Approximate Optimization Algorithm to solve the Vehicle Routing Problem. In past, QAOA has been widely used in solving various combinatorial hard optimization problems \cite{bib_Wang, bib_Gutmann, bib_Cedric}. However, looking at the results from Exp. ($\ref{exp:1}$), ($\ref{exp:2}$), and ($\ref{exp:3}$), we conclude that in general, for a finite value of $p$, there is no guarantee that the solution achieved by QAOA corresponds to the most optimal solution of the original combinatorial optimization problem \cite{bib_Will}. This is because, in QAOA instead of the following the adiabatic time evolution path, we try to guess it using $p$ steps. So, the first straightforward reason could be that the chosen value $p$ does not produce a good enough guess. Then, another reason which could explain failure of QAOA at larger values of $p$, could be the emergence of new local minimums in our solution energy-landscape which traps both gradient-free and gradient-based optimizers, and make them converge prematurely.

Moreover, in the previous studies \cite{bib_Will, bib_Alam, bib_Guerr}, it has been shown that while running QAOA on near term quantum processors, noise-based errors affect both the fidelity of state: $\ket{\beta,\gamma}$, prepared by a quantum routine, and the minimized expectation value of $\hat{H}_{cost}$, i.e., $\langle \hat{H}_{cost} \rangle$. Characterizing the behaviour of noisy quantum hardware is essential in developing error correction code, and noise-resilient algorithms. Therefore, in future, we would like to do a noise analysis of QAOA for solving the problem of VRP with additional constraints, or any other combinatorial optimization problem.

\section*{Data Availability \label{qnm_Sec5}}
The code created to run these simulations and related supplementary data could be made available to any reader upon reasonable request.

\section*{Acknowledgments}
\label{qlock_acknowledgments}
U. would like to thank IISER Kolkata and Bikash's Quantum (OPC) Pvt. Ltd. for providing hospitality during the course of the project work. B.K.B. acknowledges the support of Institute fellowship provided by IISER Kolkata.

\end{document}